\documentclass[12pt,epsf]{article}
\usepackage{graphicx}
\begin{document}

\def\a{\alpha}
\def\b{\beta}
\def\c{\varepsilon}
\def\d{\delta}
\def\e{\epsilon}
\def\f{\phi}
\def\g{\gamma}
\def\h{\theta}
\def\k{\kappa}
\def\l{\lambda}
\def\m{\mu}
\def\n{\nu}
\def\p{\psi}
\def\q{\partial}
\def\r{\rho}
\def\s{\sigma}
\def\t{\tau}
\def\u{\upsilon}
\def\v{\varphi}
\def\w{\omega}
\def\x{\xi}
\def\y{\eta}
\def\z{\zeta}
\def\D{\Delta}
\def\G{\Gamma}
\def\H{\Theta}
\def\L{\Lambda}
\def\F{\Phi}
\def\P{\Psi}
\def\S{\Sigma}

\def\o{\over}
\newcommand{\gsim}{ \mathop{}_{\textstyle \sim}^{\textstyle >} }
\newcommand{\lsim}{ \mathop{}_{\textstyle \sim}^{\textstyle <} }
\newcommand{\vev}[1]{ \left\langle {#1} \right\rangle }
\newcommand{\bra}[1]{ \langle {#1} | }
\newcommand{\ket}[1]{ | {#1} \rangle }
\newcommand{\EV}{ {\rm eV} }
\newcommand{\KEV}{ {\rm keV} }
\newcommand{\MEV}{ {\rm MeV} }
\newcommand{\GEV}{ {\rm GeV} }
\newcommand{\TEV}{ {\rm TeV} }
\def\diag{\mathop{\rm diag}\nolimits}
\def\Spin{\mathop{\rm Spin}}
\def\SO{\mathop{\rm SO}}
\def\O{\mathop{\rm O}}
\def\SU{\mathop{\rm SU}}
\def\U{\mathop{\rm U}}
\def\Sp{\mathop{\rm Sp}}
\def\SL{\mathop{\rm SL}}
\def\tr{\mathop{\rm tr}}

\def\IJMP{Int.~J.~Mod.~Phys. }
\def\MPL{Mod.~Phys.~Lett. }
\def\NP{Nucl.~Phys. }
\def\PL{Phys.~Lett. }
\def\PR{Phys.~Rev. }
\def\PRL{Phys.~Rev.~Lett. }
\def\PTP{Prog.~Theor.~Phys. }
\def\ZP{Z.~Phys. }

\newcommand{\bear}{\begin{array}}  \newcommand{\eear}{\end{array}}
\newcommand{\bea}{\begin{eqnarray}}  \newcommand{\eea}{\end{eqnarray}}
\newcommand{\beq}{\begin{equation}}  \newcommand{\eeq}{\end{equation}}
\newcommand{\bef}{\begin{figure}}  \newcommand{\eef}{\end{figure}}
\newcommand{\bec}{\begin{center}}  \newcommand{\eec}{\end{center}}
\newcommand{\non}{\nonumber}  \newcommand{\eqn}[1]{\beq {#1}\eeq}
\newcommand{\la}{\left\langle} \newcommand{\ra}{\right\rangle}

\def\SEC#1{Sec.~\ref{#1}}
\def\FIG#1{Fig.~\ref{#1}}
\def\EQ#1{Eq.~(\ref{#1})}
\def\EQS#1{Eqs.~(\ref{#1})}
\def\lrf#1#2{ \left(\frac{#1}{#2}\right)}
\def\lrfp#1#2#3{ \left(\frac{#1}{#2}\right)^{#3}}
\def\GEV#1{10^{#1}{\rm\,GeV}}
\def\MEV#1{10^{#1}{\rm\,MeV}}
\def\KEV#1{10^{#1}{\rm\,keV}}


\baselineskip 0.7cm

\begin{titlepage}

\begin{flushright}
IPMU 09-0020\\
UT-09-03
\end{flushright}

\vskip 1.35cm
\begin{center}
{\large \bf
Runaway Dynamics and Supersymmetry Breaking
}
\vskip 1.2cm
Izawa K.-I.$^{1,2}$, Fuminobu Takahashi$^{2}$, T.T. Yanagida$^{2,3}$, and Kazuya Yonekura$^{3}$
\vskip 0.4cm
{\it  
${}^1$Yukawa Institute for Theoretical Physics, Kyoto University,\\
Kyoto 606-8502, Japan\\
${}^2$ Institute for the Physics and Mathematics of the Universe,
  University of Tokyo,\\ Chiba 277-8568, Japan\\
${}^3$
Department of Physics, University of Tokyo,\\
    Tokyo 113-0033, Japan}

\vskip 1.5cm

\abstract{
Supersymmetric $SU(N_C)$ gauge theories possess runaway-type superpotentials 
for $N_F < N_C$, where $N_F$ is the flavor number of massless quarks.
We show that the runaway
behavior can be stabilized for $N_F \simeq N_C$
by introducing singlets with the aid of 
perturbative corrections to the K\"ahler potential, generating (local)
minima of supersymmetry breaking. 
 
}
\end{center}
\end{titlepage}

\setcounter{page}{2}

\section{Introduction}

It is well known that in a supersymmetric (SUSY) QCD based on an $SU(N_C)$ gauge theory with
$N_F$ flavors of quarks $Q$ and antiquarks $\tilde{Q}$~\cite{Intriligator:1995au},
the dynamically generated superpotential implies
a runaway behavior for $N_F < N_C$. In this theory
we have SUSY-invariant vacua in the limit of meson fields $|Q{\tilde Q}| \rightarrow \infty$.
This is consistent with Witten index argument for
unbroken SUSY \cite{Witten,Affleck}. In particular,
the Witten index \cite{Witten} correctly counts the number
of SUSY vacua that is determined
by adding the mass terms for all the pairs of quarks and antiquarks.

However, the situation is changed if we introduce a singlet field $S$ and assume a tree level 
superpotential of the form $SQ\tilde{Q}$. This is because
one pair of quark and antiquark is always taken massless by shifting the $S$ field even
if we introduce the mass terms for all the pairs of quarks and antiquarks, and the Witten index 
argument based on the mass deformation does not apply.
It seems that this theory can have a SUSY breaking vacuum
due to gauge dynamics in principle.
Unfortunately, we see, after integrating out quark fields, that we also
have a runaway-type 
superpotential of the singlet $S$.
Thus, it seems likely that the theory has a runaway potential for the $S$ boson and
the vacuum expectation value (vev) of $S$ goes to infinity in SUSY vacua.
Owing to this runaway behavior, this theory was not considered as a SUSY breaking model%
\footnote{We cannot exclude the possibility that there are SUSY breaking local 
minima at small vevs of the singlet field $S$ \cite{ITYY}.}.

We here note that there is subtlety in connection with quantum corrections to the K\"ahler potential.
To compute a potential in a supersymmetric theory,
we have to know both the superpotential and the K\"ahler potential of the theory. That is, 
even if the superpotential is of the runaway type,
there is a possibility that the runaway of the singlet $S$ is stopped
through the effects of the K\"ahler potential.
If such a stabilization occurs, then the $S$ has a nonvanishing $F$-term at the potential minimum
and SUSY is broken.

In this paper, we show that such a stabilization does indeed occur in a certain parameter region. 
We restrict our discussion in a weak coupling regime so that the K\"ahler potential
is calculable perturbatively. We find out that the K\"ahler potential can stop the runaway of 
the potential in a class of models with $N_F \simeq N_C \gg 1$. 
As we will see below, this stabilization is due to the anomalous
dimension of the $S$ field, which is determined mainly by 
the Yukawa coupling of the theory at the leading order
in perturbation theory.

\section{Runaway potential in SUSY QCD} \label{sec:runaway}

Let us consider a SUSY $SU(N_C)$ gauge theory with $N_F$ flavors of quarks $Q^i$ and antiquarks
${\tilde Q_i}$ ($i=1,2,\cdots,N_F$),
which belong to the fundamental and the anti-fundamental representations of
the $SU(N_C)$, respectively. 
We restrict our discussion in the case of $N_F < N_C$. We introduce a singlet chiral multiplet
$S$ and assume a tree-level superpotential
\beq
W=\lambda SQ^i{\tilde Q}_i. \label{eq:yukawa1}
\eeq
Here, we have omitted the $SU(N_C)$ indices.

Let us consider a region where the singlet $S$ has a large vev.
The vev of $S$ gives the mass $\l S$ to the quarks $Q$ and antiquarks ${\tilde Q} $ 
through the superpotential (\ref{eq:yukawa1}).
When we integrate out the massive quarks $Q$ and ${\tilde Q} $, the low energy gauge dynamics is described by
a pure $SU(N_C)$ gauge theory with the dynamical scale $\L_L$ given by%
\footnote{For the coefficients of scale matching and dynamically generated 
superpotentials, we follow Ref.\cite{Finnell:1995dr}.}
\bea
\L_L^{3N_C} = (\l S)^{N_F} \L^{3N_C-N_F},
\eea
where $\L$ denotes the dynamical scale of the high energy theory. 

In the pure $SU(N_C)$ gauge theory, the gaugino condensation occurs, which generates 
a superpotential
\bea
W_{\rm eff} = N_C \L_L^3 = N_C \L^{3-R} (\l S)^{R}, \label{eq:effsuperpot}
\eea
where $R$ denotes the ratio $R=N_F/N_C <1$.
Thus, if the K\"ahler potential took a tree level form $K=|S|^2$, the
potential of $S$ would be given by
\beq
V=N_F^2 \left|\L^{3-R} \l^{R} S^{-1+R}\right|^2, \label{eq:potential1}
\eeq 
which exhibits a runaway behavior with the vev of $S$ going to infinity.

In reality, we should take into account quantum corrections to the K\"ahler potential of $S$, 
which can change the above mentioned runaway behavior
(as well as the behavior around a small vev of the singlet field $S$,
which we do not investigate in this paper).
The low energy effective K\"ahler potential is given by
$K_{\rm eff}(|S|, M)$,
where we adopt a renormalization scheme \cite{Coleman}
\beq
\ln { \q^2 K_{\rm eff} \o \q S \q S^* } = \int_{\mu=|S|}^{M} {\tilde \g}_S(\mu) \, d (\ln \mu),
\eeq
with $M$ the renormalization point
and ${\tilde \g}_S(\mu)$ the corresponding anomalous dimension at the scale $\mu$.
The potential is obtained as
\bea
V &=& N_F^2 \left|\L^{3-R} \l^{R} S^{-1+R}\right|^2
 \left( { \q^2 K_{\rm eff} \o \q S \q S^* } \right)^{-1} \nonumber \\
 &=& N_F^2 \left|\L^{3-R} \l^{R} S^{-1+R}\right|^2 \,
 \exp \int^{|S|}_{\mu=M} {\tilde \g}_S(\mu) \, d (\ln \mu). \label{eq:easypot}
\eea

In the next section, we show by a perturbative calculation that the runaway behavior of the potential
can indeed be stopped in the theory considered above.
We see that the
$S$ is stabilized with its large vev, where the perturbative
corrections to the K\"ahler potential are estimated reliably to dominate.

\section{Stopping the runaway}

Let us consider the theory in the previous section
and show that the runaway behavior of the potential
can be stopped by the effects of the K\"ahler potential. 
We assume that the gauge coupling
is so small at the relevant energy scale $M$ that 
the perturbative calculations are reliable. 
We also assume that $1-N_F/N_C=1-R$ is very small,
which can be realized for sufficiently large $N_F \simeq N_C$.

The one-loop corrected K{\" a}hler potential is given by 
\beq
K_{\rm eff} \simeq \left( 1 - \frac{N_C N_F |\l|^2}{16\pi^2}
\ln \left|\frac{S}{eM} \right|^2 \right) |S|^2,
\eeq
which indicates ${\tilde \g}_S \simeq N_C N_F |\l|^2 /8\pi^2$.
The above one-loop result is reliable for $|S| \sim M$
with the running coupling $\l(M)$ small enough,
which enables us to ignore the higher-order corrections.

Using this K\"ahler potential and the superpotential (\ref{eq:effsuperpot}),
the potential of $S$ is determined to be
\bea
V \simeq \left( 1 + \frac{N_C N_F |\l|^2}{16\pi^2}\ln \left| \frac{S}{M} \right|^2 \right) N_F^2 \left|\L^{3-R} \l^{R} S^{-1+R}\right|^2.
\eea
Differentiating this potential with respect to $\ln |S|$,
and taking only the leading terms in $|\l|^2$ and $1-R$, we obtain
\beq
\frac{\q V}{\q \ln |S|} \simeq
 2\left(\frac{N_C N_F |\l|^2}{16\pi^2} -1+\frac{N_F}{N_C} \right)
 N_F^2 \left|\L^{3-R} \l^{R}S^{-1+R}\right|^2.
\eeq

The above perturbative results are reliable for $M \sim |S|$.
Thus we conclude that the potential has SUSY breaking minima for
\beq
{\tilde \g}_S(|S|) \simeq \frac{N_C N_F |{\tilde \l}|^2}{8\pi^2}
 \simeq 2\left(1-\frac{N_F}{N_C} \right), \label{eq:appcond}
\eeq 
with the aid of the running coupling ${\tilde \l}=\l(|S|)$.
Note that $N_F \simeq N_C \gg 1$ must be imposed for the coupling ${\tilde \l}$
to be adequately small.
The appearance of the anomalous dimension ${\tilde \g}_S$ is no accident.
In fact, this result can be directly derived
by means of the expression (\ref{eq:easypot}).

The behavior of the potential around the SUSY breaking minima can be seen
through that of the couplings.
We assume that the gauge coupling is negligibly small compared with 
the Yukawa coupling $\l$ at high energy.
Then, the renormalization group equation of $\l$ is given by
\bea
 M\frac{\q}{\q M}|\l|^2
 \simeq \frac{N_C N_F+2}{8\pi^2} |\l|^4, \label{eq:runningyukawa}
\eea
so that $|\l|^2$ as well as ${\tilde \g}_S$
is an increasing function of $M$.   
This confirms that the potential indeed has the above mentioned minima
if we take the coupling $\l$ to satisfy
\beq
\frac{N_C N_F |\l|^2}{8\pi^2} \ll 2\left(1-\frac{N_F}{N_C} \right),
\eeq
at a certain high energy scale.

Finally, a few comments are in order.
Eq.(\ref{eq:runningyukawa}) further implies that $\l$ eventually hits
a so-called Landau pole at higher energy,
which reveals that our perturbative description does not apply
to the regime of much larger $|S|$.
We also note that we still have SUSY vacua of mesonic runaway
directions with only one singlet $S$,
though such vacua may be eliminated by means of
sufficiently many singlets as in Ref.\cite{Izawa:1996pk}.


\section*{Acknowledgements}

This work was supported by the Grant-in-Aid for Yukawa International
Program for Quark-Hadron Sciences, the Grant-in-Aid
for the Global COE Program "The Next Generation of Physics,
Spun from Universality and Emergence", and
World Premier International Research Center Initiative
(WPI Initiative), MEXT, Japan.


\begin{thebibliography}{99}

%
\bibitem{Intriligator:1995au}
For a review,
  K.A.~Intriligator and N.~Seiberg,
  Nucl.\ Phys.\ Proc.\ Suppl.\  {\bf 45BC}, 1 (1996)
  [arXiv:hep-th/9509066].
%
\bibitem{Witten} E. Witten,
  Nucl.\ Phys.\  {\bf B202}, 253 (1982).
%
\bibitem{Affleck}
  I.~Affleck, M.~Dine, and N.~Seiberg,
  Nucl.\ Phys.\  {\bf B241}, 493 (1984);
  {\bf B256}, 557 (1985).
%
\bibitem{ITYY} Izawa~K.-I., F.~Takahashi, T.T.~Yanagida, and
 K.~Yonekura, arXiv:0810.5413 [hep-ph].
%
\bibitem{Finnell:1995dr}
  D.~Finnell and P.~Pouliot,
  Nucl.\ Phys.\  {\bf B453}, 225 (1995)
  [arXiv:hep-th/9503115].
%
\bibitem{Coleman}
  S.~Coleman and E.~Weinberg, Phys.\ Rev.\ {\bf D7}, 1888 (1973).
%
\bibitem{Izawa:1996pk}
  Izawa~K.-I.~and T.~Yanagida,
  Prog.\ Theor.\ Phys.\  {\bf 95}, 829 (1996)
  [arXiv:hep-th/9602180]; \\
  K.A.~Intriligator and S.D.~Thomas,
  Nucl.\ Phys.\  {\bf B473}, 121 (1996)
  [arXiv:hep-th/9603158].
%

\end{thebibliography}
\end{document}